\documentclass{article}
\usepackage{arxiv}
\usepackage[utf8]{inputenc} % allow utf-8 input
\usepackage[T1]{fontenc}    % use 8-bit T1 fonts
\usepackage{hyperref}       % hyperlinks
\usepackage{amsmath}
\usepackage{amssymb}
\usepackage{amsfonts}       % blackboard math symbols
\usepackage{float}
\usepackage{algorithm2e}
\usepackage{algorithmicx}
\usepackage{algpseudocode}
\usepackage{cancel}
\usepackage{esint}
\usepackage{mathdots}
\usepackage{mathtools}
\usepackage{mhchem}
\usepackage{stackrel}
\usepackage{stmaryrd}
\usepackage{undertilde}
\usepackage{cleveref}
\usepackage{url}            % simple URL typesetting
\usepackage{booktabs}       % professional-quality tables
\usepackage{nicefrac}       % compact symbols for 1/2, etc.
\usepackage{microtype}      % microtypography
\usepackage{lipsum}		% Can be removed after putting your text content
\usepackage{graphicx}
\usepackage[table,xcdraw]{xcolor}
\usepackage{here}
\usepackage[numbers]{natbib}
\usepackage{usebib}
\usepackage{ulem}

\title{Support-vector-machine with Bayesian optimization for lithofacies classification using elastic properties}

\author{ \href{https://orcid.org/0000-0003-0574-5311}{\includegraphics[scale=0.06]{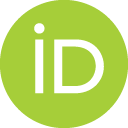}\hspace{1mm}Yohei~Nishitsuji} \\
	Department of Geoscience and Engineering\\
	Delft University of Technology\\
	Delft, The Netherlands\\
	\texttt{y.nishitsuji@tudelft.nl} \\
	\texttt{yohei.nishitsuji@gmail.com}\\
	\And
	{Jalil~Nasseri} \\
	Summit Exploration and Production Limited\\
	London, UK\\
	\texttt{jnasseri@summiteandp.com} \\
	\texttt{jalil.nasseri@gmail.com}\\
}

\begin{document}
\maketitle

\begin{abstract}
	We investigate an applicability of Bayesian-optimization (BO) to optimize hyperparameters associated with support-vector-machine (SVM) in order to classify facies using elastic properties derived from well data in the East Central Graben, UKCS. The cross-plot products of the field dataset appear to be successfully classified with non-linear boundaries. Although there are a few factors to be predetermined in the BO scheme such as an iteration number to deal with a trade-off between the prediction accuracy and the computational cost, this approach effectively reduces possible human subjectivity connected to the architecture of the SVM. Our proposed workflow might be beneficial in resource-exploration and development in terms of subsurface objective technical evaluations.
\end{abstract}

% keywords can be removed
\keywords{Deep learning \and Bayesian optimization\and Hyperparameter \and SVM \and Oil\&Gas \and Elastic properties cross-plots \and Quantitative interpretation}

\section{Introduction}
Quantitative interpretation (QI) is of importance for the purpose of improving insight from subsurface for exploration and development of natural resources. Among many seismic attributes, seismic inversion products are the most commonly used products for QI \citep[e.g.,][]{Buland1996, Harrison2012, Wang2022}. The elastic properties cross-plots are often used to assess different facies (e.g., sand and shale) and presence of hydrocarbon in reservoirs. When the properties are linearly separable in the original cross-plot space (acoustic impedance, $AI$ and velocity ratio, $\nicefrac{V_{P}}{V_{S}}$), a linear projection can be useful because of its simple and fast calculations to distinguish facies. However, the cross-plot data in practice are often exhibited as complex
clusters, which essentially requires non-linear separations. To tackle this problem, we come up with support-vector-machines \citep[SVM; e.g.,][]{Vapnik1963, Vapnik1995}, a machine-learning algorithm.

The SVM is originally developed to solve a binary (two) classification problem by a linear separation. Having called kernel-function, however, the SVM can handle non-linear problems too. Moreover, not only the binary
problems but also multiclass-classification problems can be solved when an appropriate aggregation strategy is implemented. A couple of hyperparameters require to be optimized to perform the SVM effectively and efficiently for any given problems. In terms of finding optimal hyperparameters, the simplest approach can be an exhaustive-grid search \citep[e.g.,][]{Wang2014}. This method seeks every single grid within a given searching range, literally one by one. Even though when the number of the SVM's hyperparameters to be optimized is small, this method becomes an inefficient approach when the searching range becomes huge. A time-efficient approach we would like to apply in this study is called Bayesian-optimization \citep[BO; e.g.,][]{Mockus1978}. 

The BO is a global optimization algorithm that finds hyperparameters of black-box (unknown) functions in as few iterations as possible by searching a good balance between so-called exploration and exploitation \citep[e.g.,][]{Desautels2014}. There are many industries and applications that come up with the BO in practice such as social network, accommodation fare metasearch and travel route aggregator. The exploration factor seeks where the variance of the current posterior measurement is high for next round of evaluation (iteration), whereas the exploitation factor focuses on where the means of the posterior is low. In other words, the exploration part completely ignores what the posterior has already obtained so far, whilst the exploitation part completely ignores what the posterior has not obtained so far.

\begin{figure}[h]
	\centering
	\includegraphics[width=100mm]{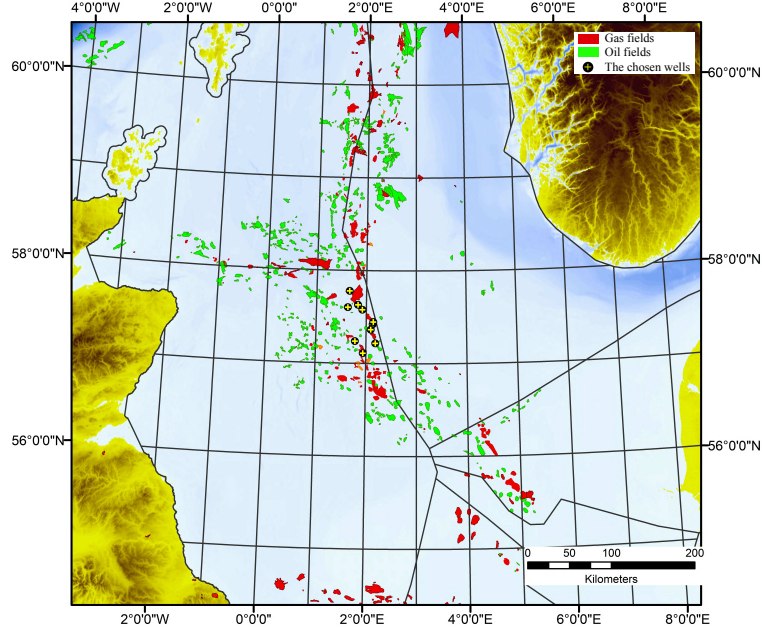}
	\caption{The chosen 10 wells (see the introduction for more details) for the study in the East Central Graben in the UK North Sea.}
	\label{fig:fig1}
\end{figure}

In this study, we utilize the BO to optimize the SVM's hyperparameters for the multiclass classification problems associated with elastic impedance cross-plot data obtained from well data in the East Central Graben in UKCS (Figure \ref{fig:fig1}). We use 10 wells in this study including 22/9-5, 22/14b-5, 22/15-3, 22/15-4, 22/24c-11, 22/30a-16, 23/16d-6, 23/16b-9, 2316f-12, and 23/21-5 (UK National Data Repository: \href{https://ndr.ogauthority.co.uk}{https://ndr.ogauthority.co.uk}). The target interval in this study is the Paleocene sediments where the Forties member of Sele Formation is the main reservoir sandstone.

\section{The Bayesian-Optimization Applied to The Support-Vector-Machine}
\subsection{Multiclass Classification of The Support-Vector-Machine}
\label{sec:headings}
\begin{figure}[h]
	\centering
	\includegraphics[width=115mm]{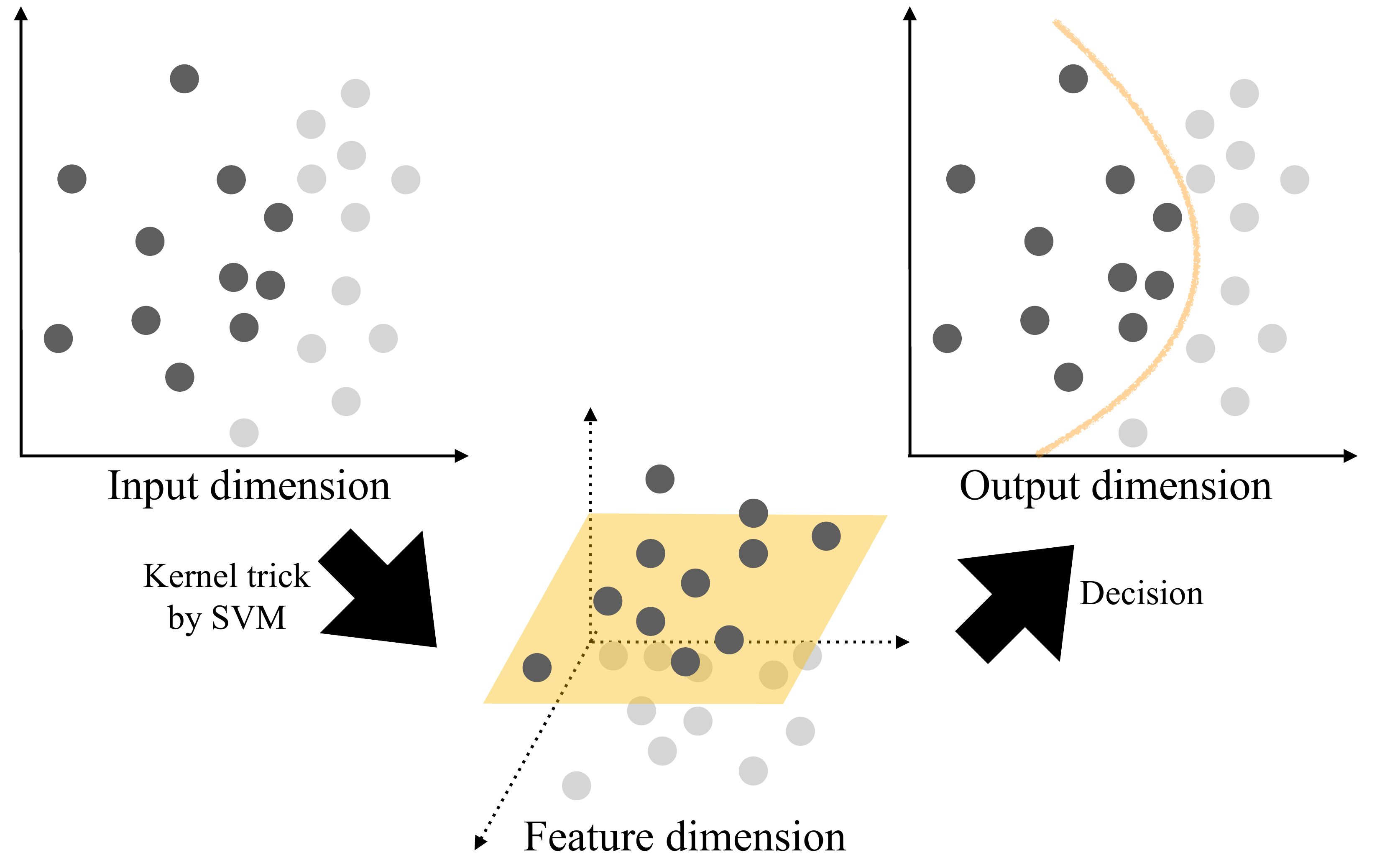}
	\caption{Schematic image how SVM classifies the binary problem non-linearly. With the kernel trick, the original input dimension is implicitly converted to the feature dimension where the decision boundary can linearly separate classes. The estimated linear boundary in the feature dimension then converts back to the original dimension being as the non-linear boundary.}
	\label{fig:fig2}
\end{figure}

In this section, we briefly introduce the multiclass classifications by the SVM. The SVM, which is based on statistical learning theory and structural risk minimization, was initially developed for the purpose of separating binary problems \citep[e.g.,][]{Vapnik1995}. We show a schematic of the binary problem using the SVM in Figure \ref{fig:fig2}. Considering of actual classifications in practice, however, there are many cases that we need to deal with more than three classes. For example, five chosen classes in this study are the geological facies such as shale, tuff, brine-, oil-, gas-sandstone. A plenty number of studies connected to aggregation strategy has been reported to deal with multiclass classifications \citep[e.g.,][]{Galar2011}. To date, this is one of ongoing-research topics in machine-learning community. 

We assume that we have a training dataset $T$ which has $p$-th classes for the multiclass classifications by the SVM, such that:

\begin{flalign}
\label{eq1}
T=\left\{ \left(\mathbf{x}_{1},y_{1}\right),\left(\mathbf{x}_{2},y_{2}\right),\cdots\left(\mathbf{x}_{n},y_{n}\right)\right\} ,\,\mathbf{x}_{i}\in\mathbb{R^{\mathrm{\mathit{m}}}},\,y_{i}\in\{1,2,\cdots,p\}, &  & {}
\end{flalign}

where $\mathbf{x}_{i}$ and $y_{i}$ denotes $i$-th feature vector whose feature dimension is $m$ and their associated classes (labels). For example, $p=2$ means a binary-class problem. The SVM minimizes a difference between the expected risk, which we cannot directly obtain due to the lack of probabilistic-distribution information including unknown data which are not observed yet, and empirical risk \citep[e.g.,][]{Vapnik1995}. The minimization allows us to find a maximum margin which divides $\mathbf{x}_{i}$ into two classes if $p=2$. The decision boundaries drawn by the margins can be either linear or non-linear via hyperplane using so-called kernel function. Among several kernel functions, the gaussian kernel is commonly used because of its better performance than the others \citep[e.g.,][]{Scholkopf1997}. Therefore, we adopt the gaussian kernel in this study.

There are two hyperparameters require optimization for a possible gain in performance of the SVM. The first parameter is called sigma, which presents in a gaussian kernel. This parameter determines the complexity (non-linearity) of the margins. The second parameter is penalization, which grants the rate of misclassification. The penalization parameter can be derived from the Lagrangian-dual-problem for the SVM (not shown here for the sake of brevity). Moreover, BO determines the type of aggregation strategy of either one-versus-one (OvO) or one-versus-all (OvA) for solving multi binary class problems.

\subsection{The Gaussian-Process}
\label{sec:headings}
We adapt the BO for the optimization framework and determine the aforementioned two hyperparameters and the aggregation strategy for optimizing SVM. With the regime of the BO, a probability distribution to an unknown-target function $f$, in which hyperparameters $h_{i}$ are present, is assumed to follow the gaussian-process prior:

\begin{flalign}
\label{eq2}
p\left(\begin{array}{c}
f(h_{1})\\
f(h_{2})\\
\vdots\\
f(h_{n})
\end{array}\right)\sim\mathcal{N}\left\{ \left(\begin{array}{c}
\mu(h_{1})\\
\mu(h_{2})\\
\vdots\\
\mu(h_{n})
\end{array}\right),\left(\begin{array}{cccc}
K(h_{1},h_{1}) & K(h_{1},h_{2}) & \cdots & K(h_{1},h_{n})\\
K(h_{2},h_{1}) & K(h_{2},h_{2}) & \cdots & K(h_{2},h_{n})\\
\vdots & \vdots & \ddots & \vdots\\
K(h_{n},h_{1}) & K(h_{n},h_{2}) & \cdots & K(h_{n},h_{n})
\end{array}\right)\right\} , &  & {}
\end{flalign}

where $\mathcal{N}$ , $\mu$ and $K$ indicate normal (gaussian) distribution, the mean function (usually set to be zero) and the covariance function which can be derived from the utilized the gaussian kernel. As one can see from equation \ref{eq2}, the gaussian process, which is completely defined by the mean function as well as the covariance function, is a group of random variables whose distributions are all based on the gaussian distributions. While the covariance is independent of observations, the mean is a linear combination of each observation. For the brevity, we simplify equation \ref{eq2} using matrix-vector notation:

\begin{flalign}
\label{eq3}
p(\mathbf{f})\sim\mathcal{N}(\mathbf{f}; & \boldsymbol{\mu},\mathbf{K}), & {}
\end{flalign}

where $\mathbf{K}$ is a $n\times n$ matrix, whilst $\mathbf{f}$ and $\boldsymbol{\mu}$ are $n\times1$ column vectors. $p(\mathbf{f})$ is the prior information that can be used in the Bayesian regression \citep[e.g.,][]{Shahriari2016}, which is given by:

\begin{flalign}
\label{eq4}
p(\mathbf{f}\mid\mathbf{D})=\frac{p(\mathbf{f})p(\mathbf{D}\mid\mathbf{f})}{p(\mathbf{D})} & , & {}
\end{flalign}

where $\mathbf{D}$ is a set of observation data (also often called evidence) in the matrix-vector notation of $D_{1:n}\left\{ h_{1:n},f(h_{1:n})\right\} $. Given new evidence, a posterior probability, the left-hand side term in equation \ref{eq4}, is updated. Note that we assume noise-free case in equation \ref{eq4} for simplicity. By solving equation \ref{eq4}, we expect to choose where to observe the function next. The following equation is adapted to make the choice to be fully automated:

\begin{flalign}
\label{eq5}
\mathbf{h}_{n+1}=\underset{\mathbf{h}}{argmax}\,\alpha(\mathbf{h};\mathbf{D}) & , & {}
\end{flalign}

where $\alpha$ is an acquisition function, which contributes to the objective function $\underset{\mathbf{h}}{argmax\,}\mathbf{f}.$ Although several policies (tasks and criteria) associated with the acquisition function have been ever proposed in the machine-learning community \citep[e.g.,][]{Kushner1964, Mockus1978, Snoek2012}, the following policy explained hereinafter is used in this work.

\subsection{The Expected-Improvement-Per-Second}
\label{sec:headings}
The expected improvement, developed by Jones et al.~\cite{Jones1998} based on the pioneer work by Mockus et al.~\cite{Mockus1978}, is commonly used within the number of the acquisition functions. The policy of this function is to evaluate the expected value of possible improvement of the objective function. A searching point where to observe next is based on the largest expected improvement. The acquisition function of the expected improvement can be written as:

\begin{flalign}
\label{eq6}
\alpha=\int_{-\infty}^{\infty}max(0,\mathbf{f}-\mathbf{f}_{best},)\mathcal{N}(\mathbf{f};\boldsymbol{\mu},\mathbf{K})d\mathbf{f} & , & {}
\end{flalign}

in which the next searching point with the largest expected improvement is automatically selected one after the other. It can be noted that the left-hand side term in the equation \ref{eq6} increases when the mean function and the covariance (Kernel) function decreases and increases, respectively. Therefore, this policy implicitly has a trade-off between the exploitation factor (driven by low mean) and the exploration factor (driven by high covariance). Tendentially, an explicit trade-off parameter exists in other policies such as the upper-confidence based policy \citep[e.g.,][]{Srinivas2010} which we do not use here.

Snoek et al.~\cite{Snoek2012} improved the performance of the expected improvement by putting the time-weight on its evaluation time, which is called the expected-improvement-per-second (EIPS). The EIPS seeks a point to evaluate the next from where the value of the expected improvement per second is largest. The EIPS finds the optimal parameters faster than using the expected improvement by the Markov-chain-Monte-Carlo \citep[e.g.,][]{Snoek2012}. In our study, we adopt the EIPS policy for the BO to automatically optimize the hyperparameters of the SVM with respect to solve the facies-classification problems associated with the elastic properties cross-plot data.

\section{Implementation Test}
\label{sec:headings}
\begin{figure}[h]
	\centering
	\includegraphics[width=90mm]{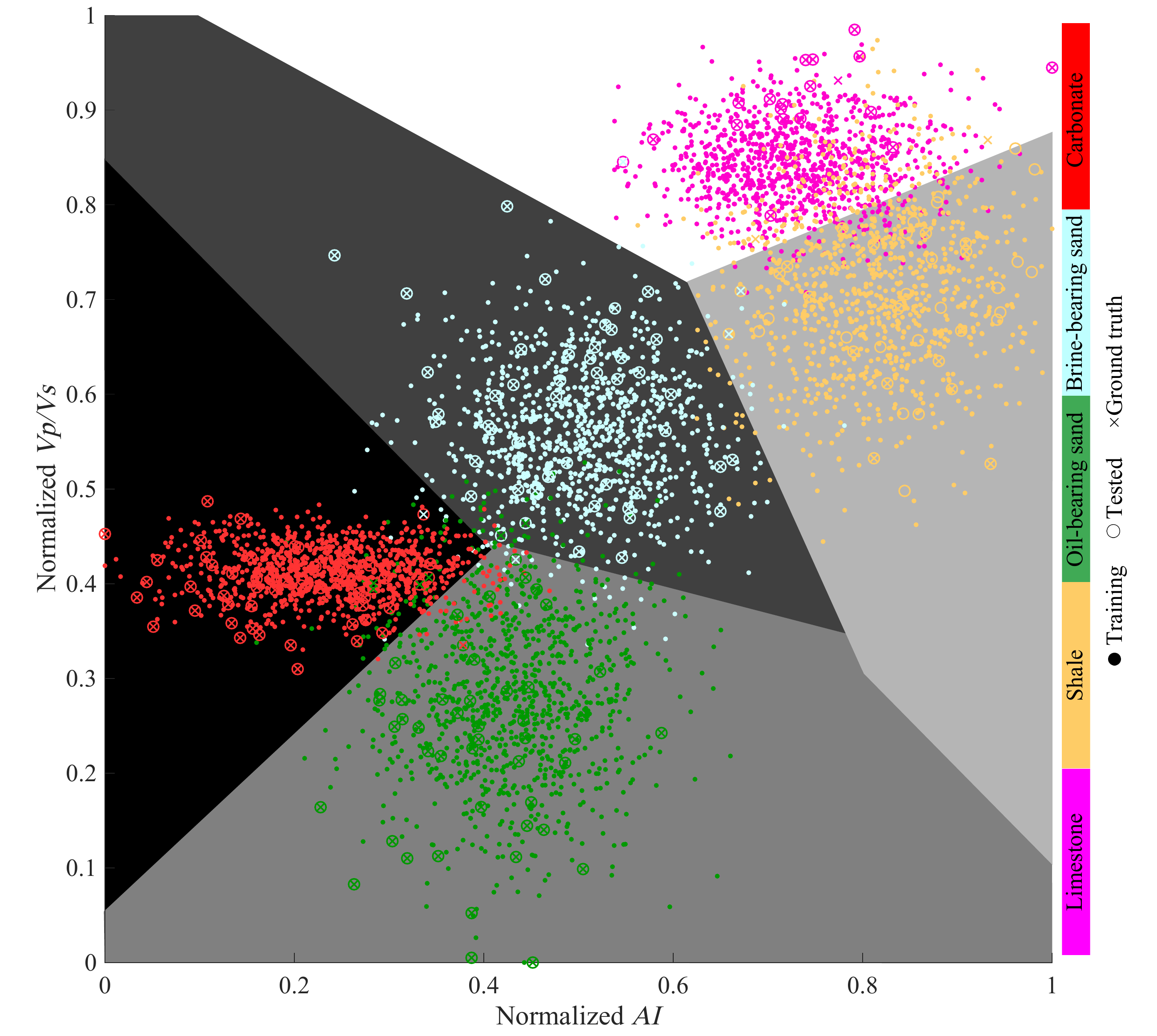}
	\caption{Classification result of synthetic elastic properties ($AI$ and $\nicefrac{V_{P}}{V_{S}}$) using the SVM after 20 BO iterations of the hyperparameters. The filled circles are the training data, the hollow circles are the test data, and the crosses are the ground truth. C, B, O, S and L in the legend box stand for condensate, brine-bearing sand, oil-bearing sand, shale and limestone facies. The five different background colours correspond with the classified facies. The details of 20 BO iterations are shown in Figure \ref{fig:fig4}, and a confusion matrix of the test accuracy is shown in Figure \ref{fig:fig5}.}
	\label{fig:fig3}
\end{figure}

We carry out a synthetic test to check feasibility of our computational implementation on an $AI$-$\nicefrac{V_{P}}{V_{S}}$ cross-plot prior to the actual classification using the field dataset. By means of the feasibility test we assure that there are no technical problems caused by our implementation per se. The number of target facies defined by the well data for classification is five in this study. We therefore synthesize five classes. We generate 1,000 supervisors (training data) and 50 test data for each class using a gaussian randomizer. After training the SVM with the BO, we check the performance of the test classification. The learning process in general can be optimized within 20 iterations by the BO \citep{Czarnecki2015}, so we execute 20 iterations. The strategy of the multiclass classification is either the OvO or the OvA depending on the BO. The classification results are shown in Figure \ref{fig:fig3} using the EIPS acquisition function.

\begin{figure}[h]
	\centering
	\includegraphics[width=120mm]{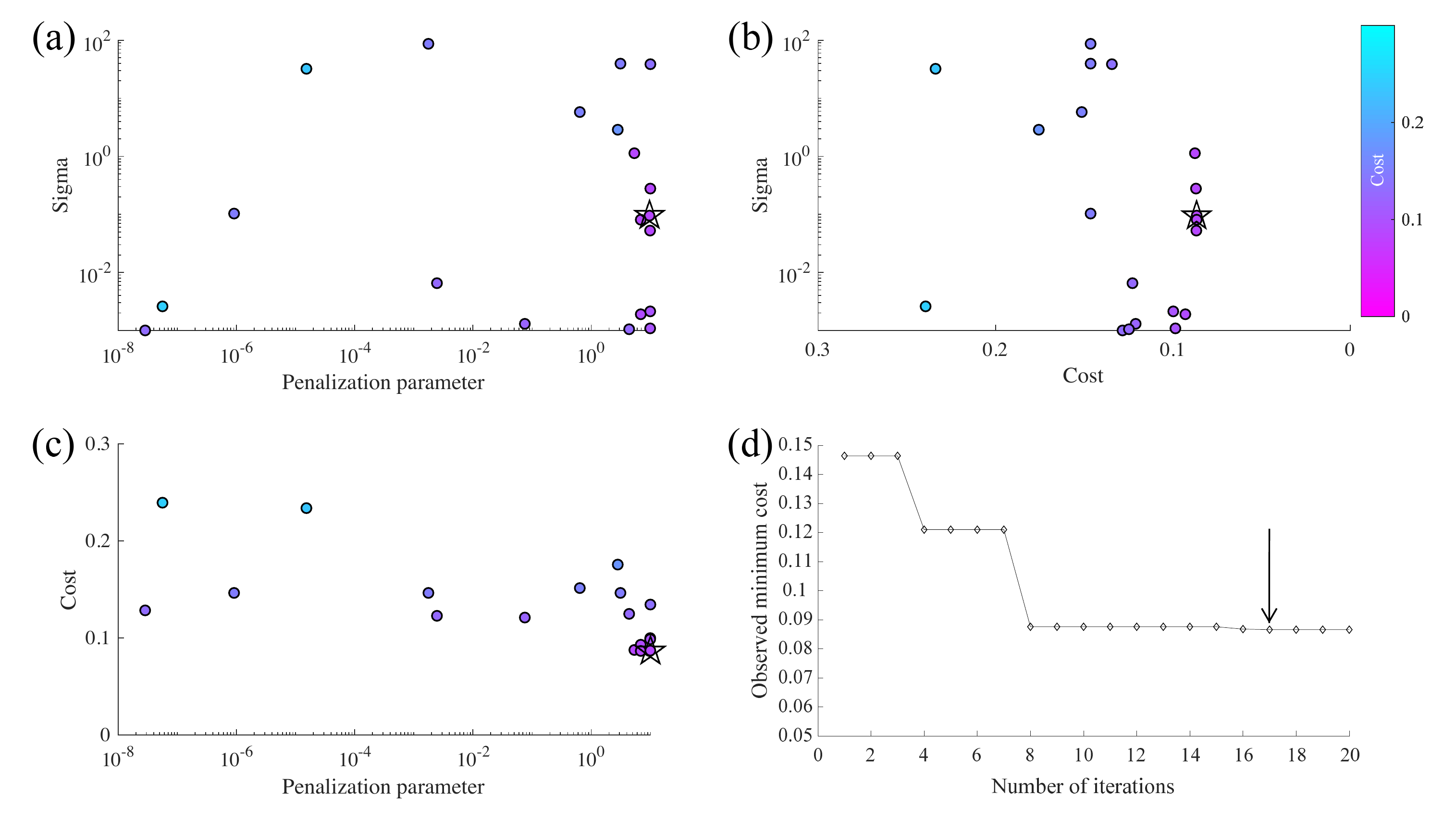}
	\caption{Cross plots of cost values after 20th BO iteration using the synthetic data (Figure \ref{fig:fig3}) for: (a) the penalization parameter and sigma; (b) the cost and the sigma; (c) the penalization parameter and the cost; (d) the number of iterations and the minimum cost. The stars and arrow correspond to where the optimal values are found.}
	\label{fig:fig4}
\end{figure}

\begin{figure}[h]
	\centering
	\includegraphics[width=70mm]{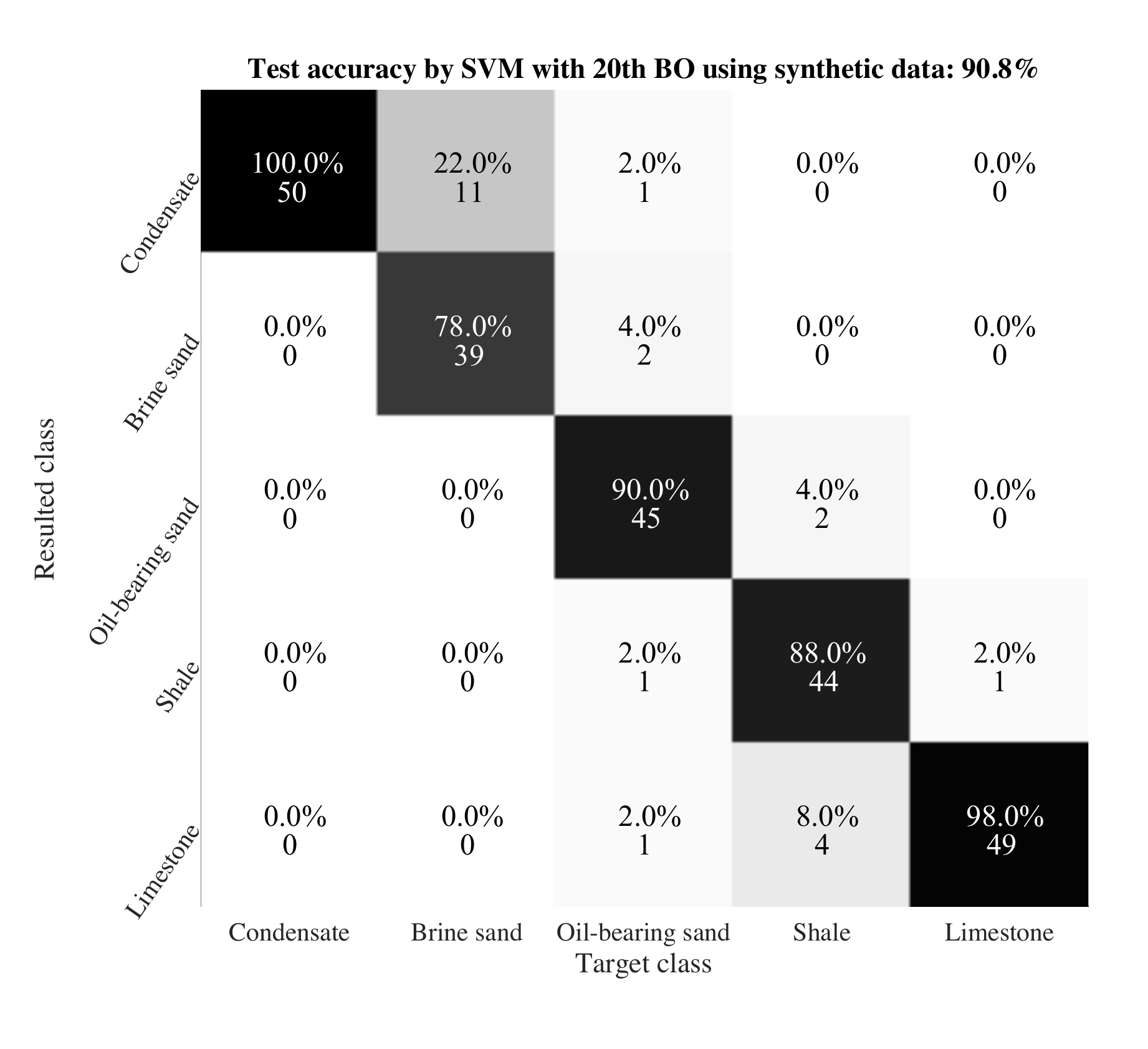}
	\caption{Confusion matrix of the synthetic data classification in Figure \ref{fig:fig3}. The percentile in each bin is the test accuracy between the target- and resulted classes. The integer shown in each bin is the used sample number.}
	\label{fig:fig5}
\end{figure}

The cost function in Figure \ref{fig:fig4} shows that the optimized hyperparameters are consequently found in 17th iteration, but major improvement of the cost is obtained in 8th iteration. From Figures \ref{fig:fig4}a-c we find out that the BO searches the hyperparameters from both aspects of the exploitation and the exploration followed by equation \ref{eq6}. This means that several iterations (represented by circles in Figure \ref{fig:fig4}a-c) are focused on neighbouring locations (e.g., where the penalization parameters are higher than about 2\textasteriskcentered 10\textsuperscript{0} in Figure \ref{fig:fig4}c) by the exploitation factor whilst some iterations are randomly realized by the exploration factor. Based on this experiment, we do not observe major technical issues caused by our implementation. In Figure \ref{fig:fig5}, we plot a confusion matrix for the test accuracy of 90.8 \%. The numbers in each bin indicate the test accuracy per target class when the sample numbers in the identical bin are used. Note that the accuracy shown here stands for the total accuracy of all kinds of used facies. In this synthetic example, two classes of oil and gas bearing sands are well separated and distinguished from the others as the main target. A prediction accuracy of 90.0 \% is achieved with only 20 iterations for the utilized BO.

\section{Results and Discussions}
\label{sec:headings}
\begin{figure}[h]
	\centering
	\includegraphics[width=90mm]{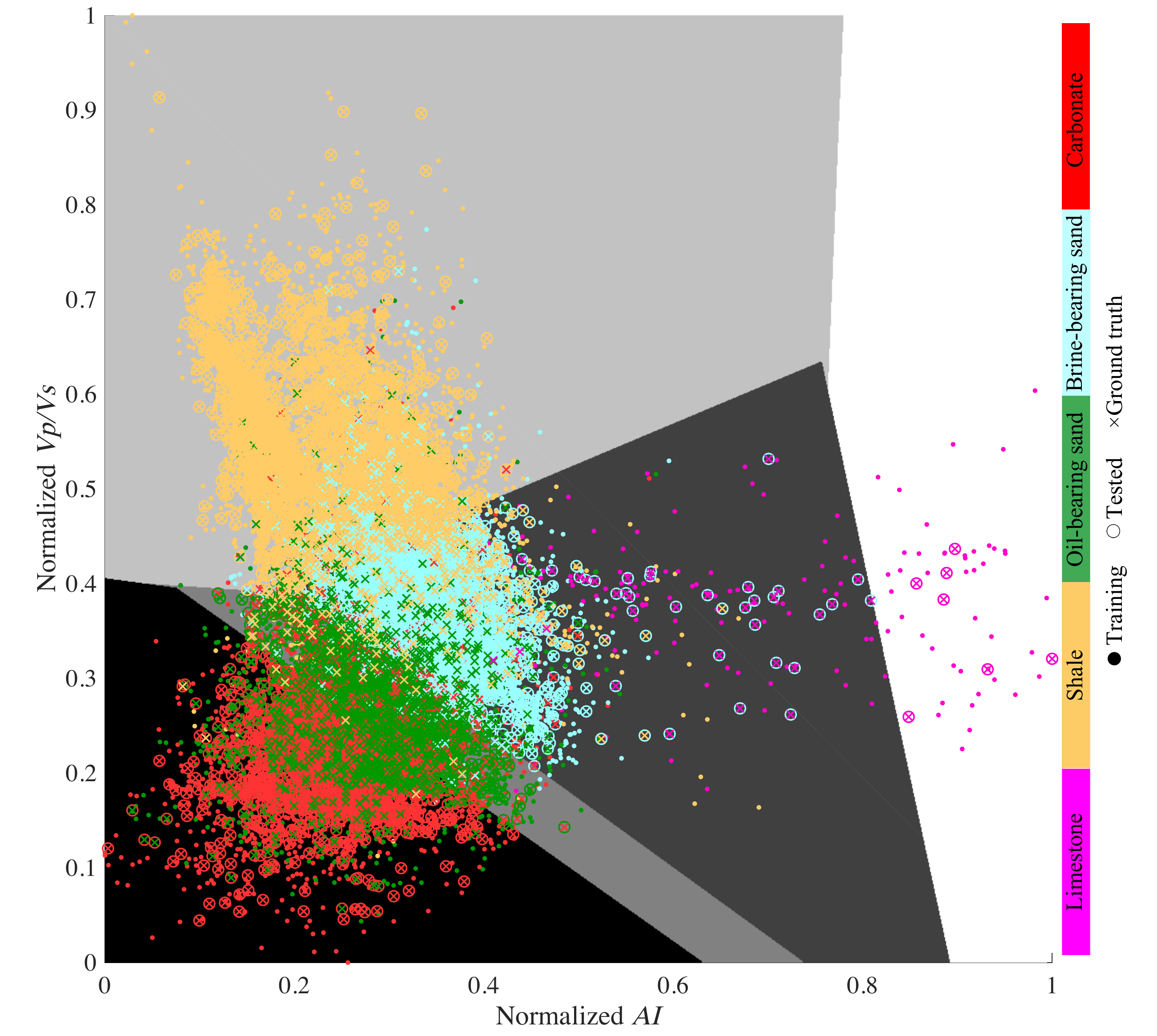}
	\caption{Facies classification result using elastic properties ($AI$ and $\nicefrac{V_{P}}{V_{S}}$) from the chosen 10 wells through the SVM after 20th BO iteration of the hyperparameters. The legend and the background colours are same as Figure \ref{fig:fig3}. The details of 20 BO iterations are shown in Figure \ref{fig:fig7}, and a confusion matrix of the test accuracy is shown in Figure \ref{fig:fig8}a.}
	\label{fig:fig6}
\end{figure}

\begin{figure}[h]
	\centering
	\includegraphics[width=120mm]{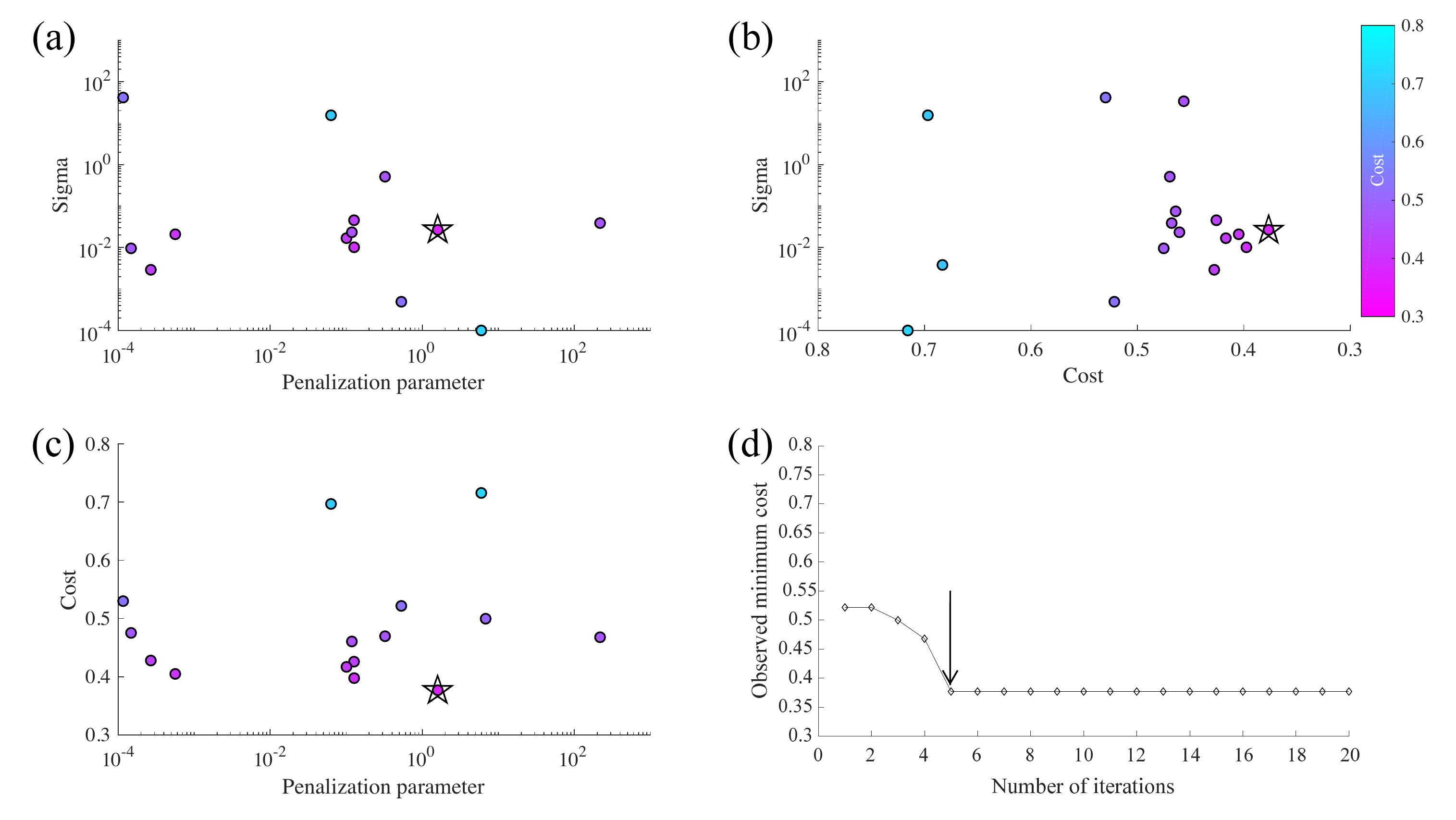}
	\caption{Same as Figure \ref{fig:fig4}, but for the field dataset (Figure \ref{fig:fig6}).}
	\label{fig:fig7}
\end{figure}

\begin{figure}[h]
	\centering
	\includegraphics[width=170mm]{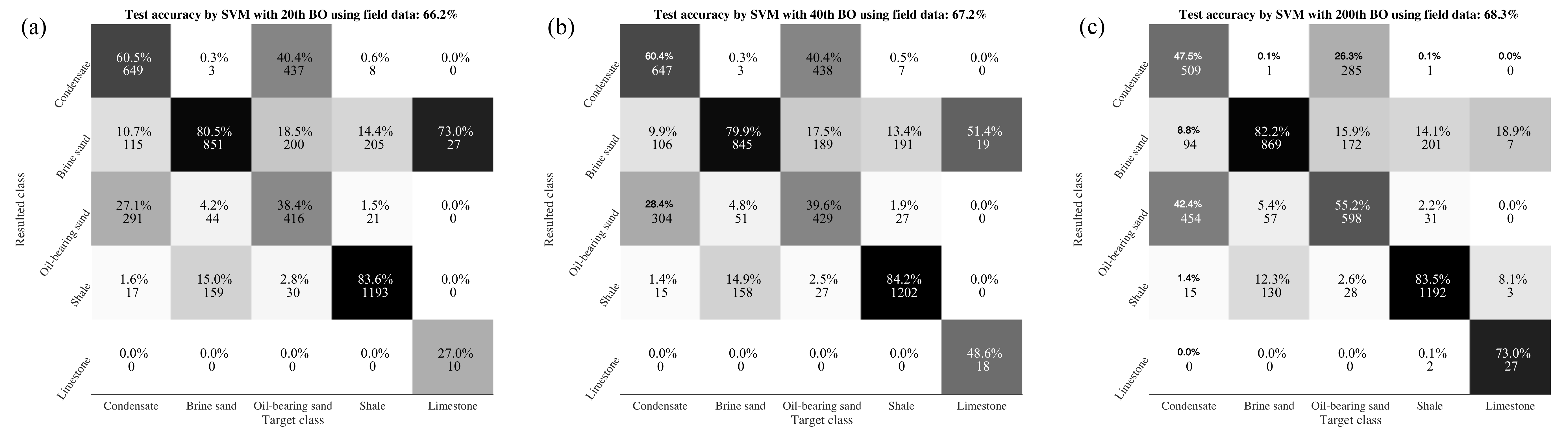}
	\caption{Confusion matrices of the field data classification with: (a) 20 iterations in Figures \ref{fig:fig6} and \ref{fig:fig7}; (b) 40 iterations in Figures \ref{fig:fig9} and  \ref{fig:fig10}; (c) 200 iterations in Figures \ref{fig:fig11} and  \ref{fig:fig12}.}
	\label{fig:fig8}
\end{figure}

The Figure \ref{fig:fig6} shows the cross-plot of the elastic parameters, $AI$ and $\nicefrac{V_{P}}{V_{S}}$, for the Palaeocene formations derived from ten chosen wells. There are always overlaps between different facies/classes and we intend to improve and maximize the separation between the classes by implementing the methods described in the sections above. The initial classification results show the optimized parameters are consequently found in 5th iteration out of 20 iterations (Figure \ref{fig:fig7}). The test (prediction) accuracy after 20 iterations result in 66.2 \% at this realization (Figure \ref{fig:fig8}a). Note that the test accuracy might differ slightly after running consequent realizations due to the dynamic nature of the BO.

The Figure \ref{fig:fig7} indicates the sigma are intensively exploited between 10\textsuperscript{0}and 10\textsuperscript{-2} in Figure \ref{fig:fig7}b, where the cost is relatively low among all of the 20 iterations. On the contrary, we also see that the BO tries to explore where there are little observations available i.e., the sigma around 10\textsuperscript{-4} in Figure \ref{fig:fig7}b. 

The decision boundaries in the initial classification in general appear to be good visually (Figure \ref{fig:fig6}), however, there are rooms to improve the separation between the facies. For instances, it looks like that an alternative boundary between the brine sand and the limestone can improve the separation rate such that the limestone class with the white background in Figure \ref{fig:fig6} could have been more expanded towards left by exchanging with the brine-sand class with the dark grey background. On the contrary, it is too hard to uphold the decision boundaries between classes with significant overlaps such as the oil- and condensate-bearing sands or brine sand and shale visually. However, any further improvement in other scenarios can be examined quantitatively by the confusion matrix, i.e., Figure \ref{fig:fig8}a. A certain number of miss-classified plots is inevitable because of faceis overlaps on $AI$-$\nicefrac{V_{P}}{V_{S}}$ cross-plot. When we come to think of such over-lapping cases in general, our subjectivity might prone to be dictated unless automatic methods like the BO-based-SVM are adopted.

\begin{figure}[h]
	\centering
	\includegraphics[width=90mm]{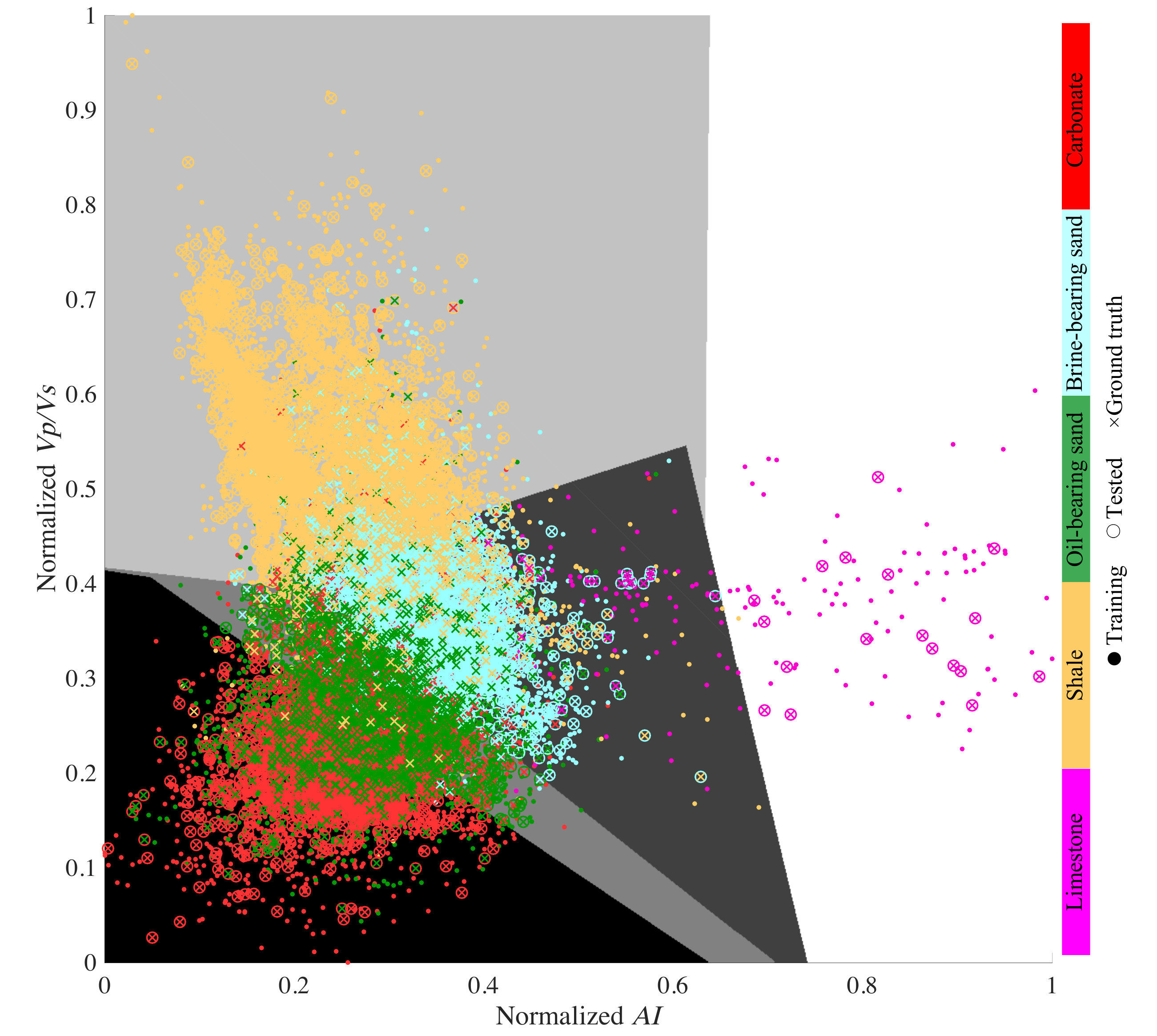}
	\caption{Same as Figure \ref{fig:fig6} but for 40 iterations. The details of 40 BO iterations are shown in Figure \ref{fig:fig10}, and a confusion matrix of the test accuracy is shown in Figure \ref{fig:fig8}b.}
	\label{fig:fig9}
\end{figure}

\begin{figure}[h]
	\centering
	\includegraphics[width=120mm]{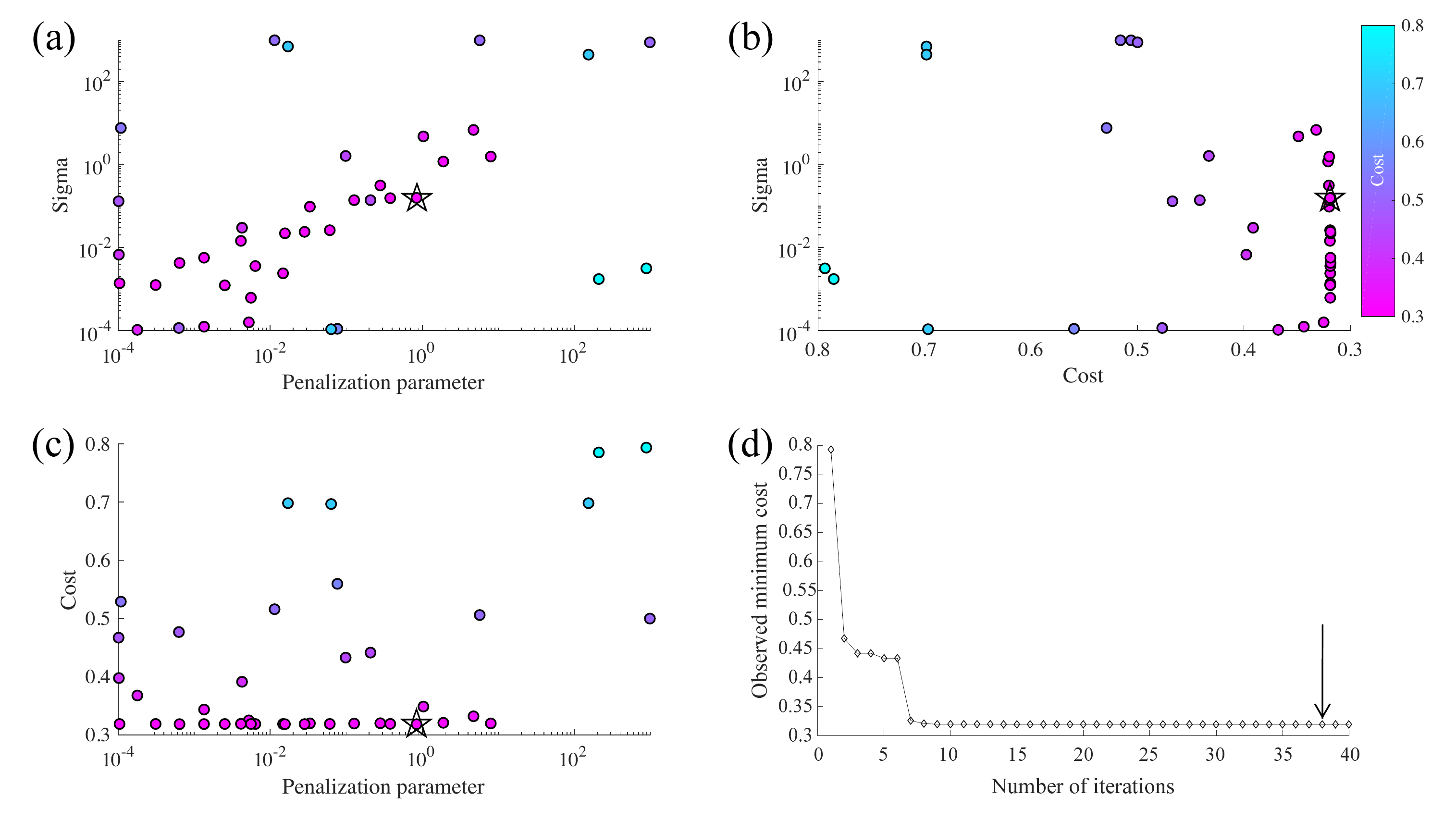}
	\caption{Same as Figure \ref{fig:fig7} but for 40 iterations (Figure \ref{fig:fig9}).}
	\label{fig:fig10}
\end{figure}

We investigate further improvement in classification and alternative decision boundaries by increasing the number of iteration; we doubled it to be 40 times. The results are shown in Figures \ref{fig:fig8}b-\ref{fig:fig10}.

Although the improvement is consequently found in the 38th iteration, the large improvement of the cost function is rapidly gained within the first several iterations (Figure \ref{fig:fig10}d). The searching range of the hyperparameters in Figure \ref{fig:fig10} gets broader for both small and large values than those in Figure \ref{fig:fig7}. A possible correlation between the sigma and the penalization parameter in terms of the cost begins to appear in Figure \ref{fig:fig10}a. The most of pairs with the cost show lower than 0.4 to be likely emerged when both the sigma and the penalization parameter have similar values in logarithmic scale. This is not surprising when we consider of the hyperparameters associated with the SVM. Recall that the sigma is responsible to the margins' non-linearity whereas the penalization parameter is related to the misclassification. Hence, it can be interpreted that the BO automatically responds to the trade-off nature of these two hyperparameters.

\begin{figure}[h]
	\centering
	\includegraphics[width=90mm]{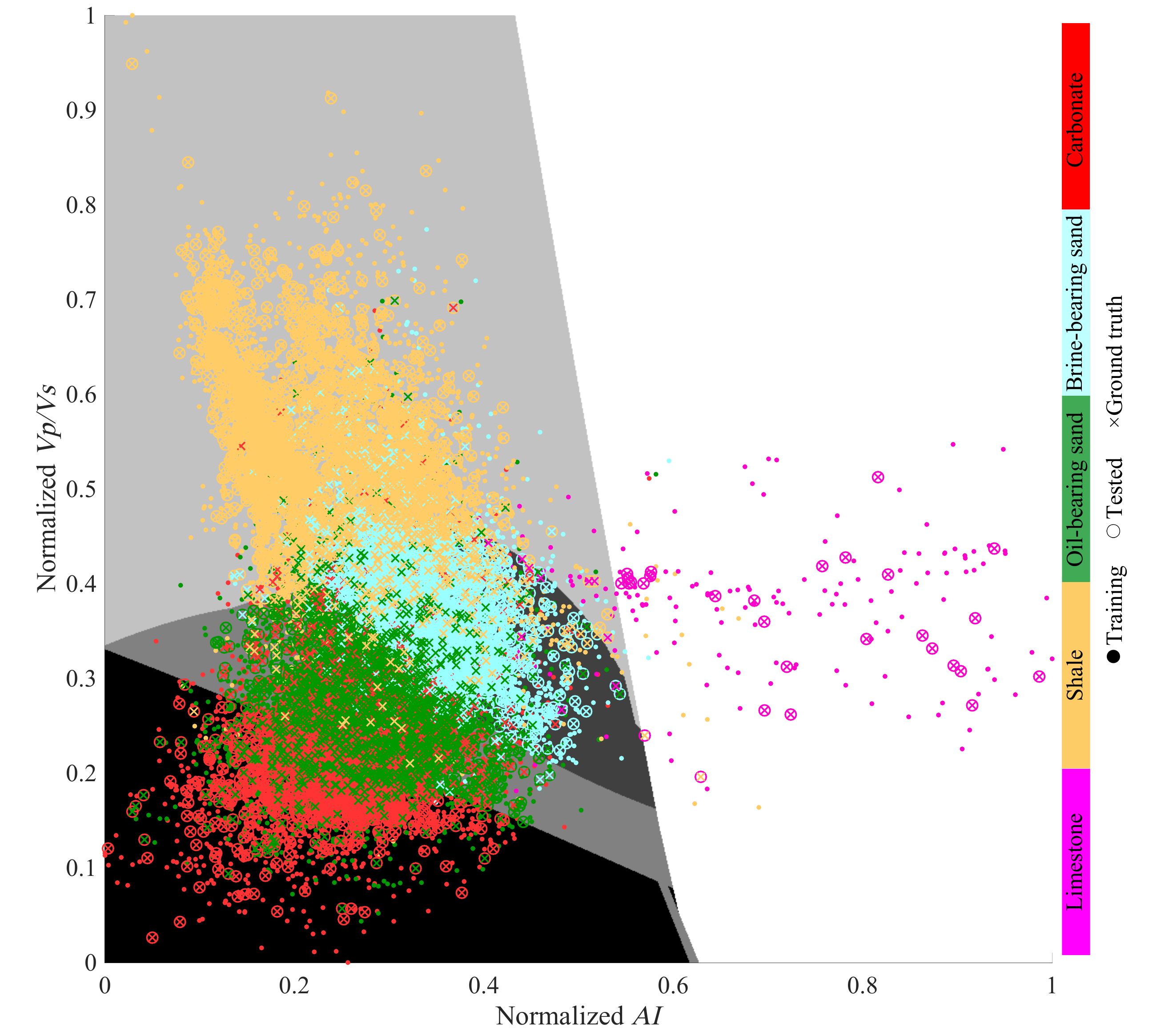}
	\caption{Same as Figure \ref{fig:fig6} but for 200 iterations. The details of 200 BO iterations are shown in Figure \ref{fig:fig12}, and a confusion matrix of the test accuracy is shown in Figure \ref{fig:fig8}c.}
	\label{fig:fig11}
\end{figure}

\begin{figure}[h]
	\centering
	\includegraphics[width=120mm]{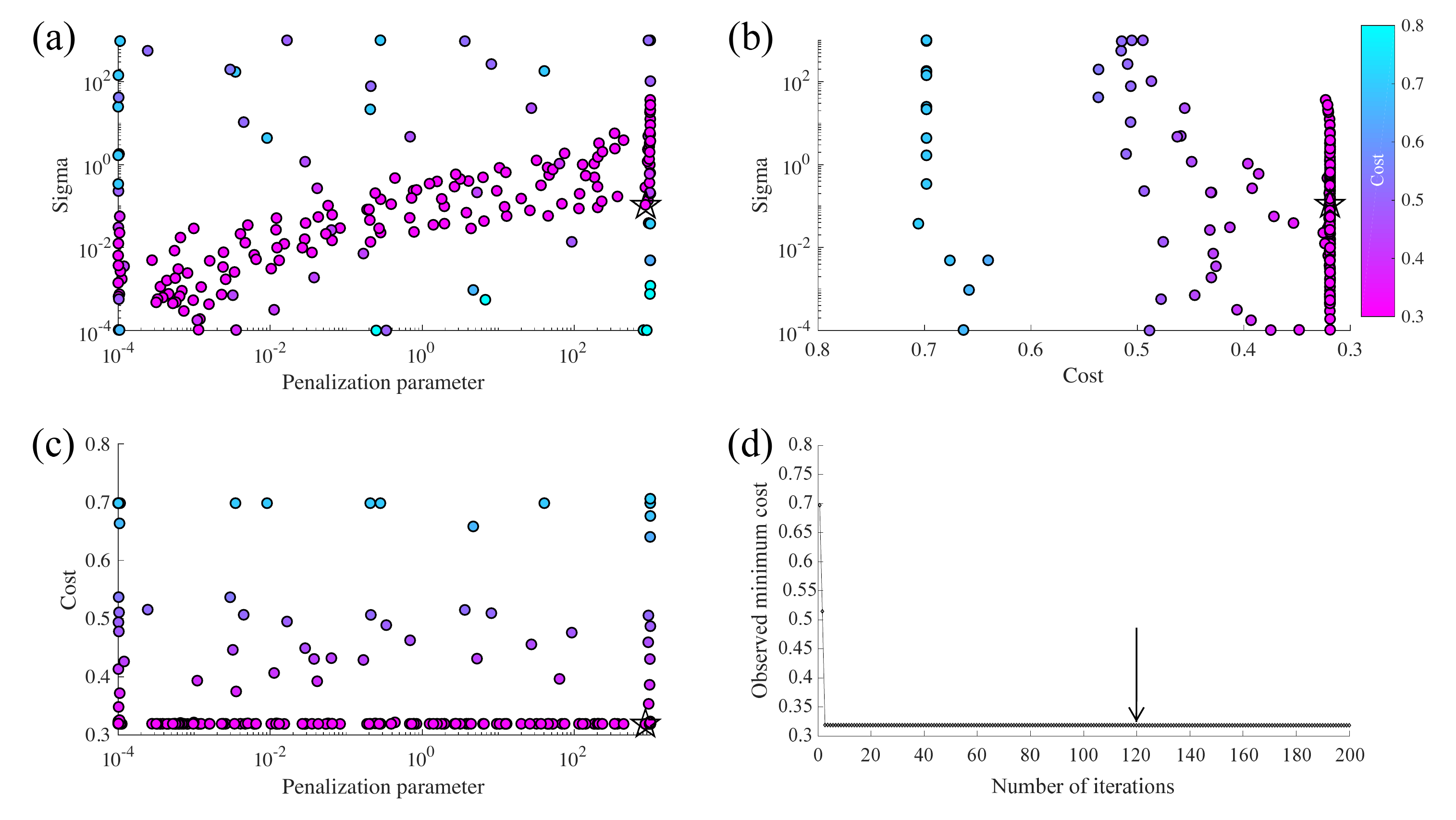}
	\caption{Same as Figure \ref{fig:fig7}, but for 200 iterations (Figure \ref{fig:fig11}).}
	\label{fig:fig12}
\end{figure}

There are certain improvements in the new classification scenario in Figure \ref{fig:fig9} in comparison with Figure \ref{fig:fig6} such as visible changes in the boundary between the brine sand and the limestone. Not only the total test accuracy but also the accuracy of the oil-bearing sand in the new classification increases by 1 \% (Figures \ref{fig:fig8}a-b). Since we still visually see the potential improvement of the aforementioned boundary between the brine sand and the limestone, we carry out one more additional test with 200 iterations. The results are shown in Figures \ref{fig:fig8}c, \ref{fig:fig11} and \ref{fig:fig12}.

Like other last two realizations by the 20- (Figure \ref{fig:fig7}d) and the 40-iteration (Figure \ref{fig:fig10}d), the 200-iteration (Figure \ref{fig:fig12}d) also shows a major improvement of the cost function within the first several BO iterations. The possible correlation discussed in Figure \ref{fig:fig10}a becomes more prominent (Figure \ref{fig:fig12}a). In overall, the 200-iteration yields more than 1 \% improvement of the test accuracy. For the oil-bearing sand, the accuracy improves to 55.2 \%, which is about 15 \% higher than when the 40-iteration is tested. 

As demonstrated hitherto, the hyperparameters of the SVM can be automatically optimized within the framework of the BO. When the cost function appears to be even stable (e.g., Figure \ref{fig:fig7}d), the test accuracy might be improved further by simply increasing the iteration if the current classification is far from the global minima. The iteration number might be determined by a trade-off between the targeted test accuracy and the computational cost. For the field dataset used in the study, there is a fundamental limitation for the classifications relating to the overlapping facies or classes. Some additional elastic properties as input might be considered in future investigations in order to achieve better classification beyond the current limitation, which also would cost more computations. With this regard, so-called multidimensional scaling \citep[e.g.,][]{Borg1997} could be useful to be implemented.

The computational cost by the grid search tends to be enormous especially when the searching range is large but utilizing the BO is a versatility choice for geophysical applications as long as we use conventional computers rather than quantum computers which are under development.

\section{Conclusions}
We presented an application of Bayesian optimization (BO) with support vector machine (SVM) in order to classify cross-plot products of elastic properties which could be used in seismic QI studies. The applications showed in this study used the data from 10 wells in East Central Graben in the UKCS. We achieved the classification results with visually feasible appearance having non-linear decision boundaries without choosing the SVM hyperparameters subjectively. The cost function related to hyperparameters optimization by the BO exhibited to be stable before finishing the first dozen iterations, but the test accuracy and the visual classification slightly improved by increasing the total iteration number. However, it is difficult to predetermine which iteration number is optimal because there is a trade-off between the prediction accuracy and the computational cost. Nonetheless, the BO application is potentially more efficient in optimizing the hyperparameters of the SVM than a conventional exhaustive-grid search. So, the application of BO with SVM can be useful for well and consequently seismic quantitative interpretations and facies analysis in resource exploration and development subsurface studies. 

\subsection*{Acknowledgments}
For SVM applications and Bayesian optimization, LIBSVM and BayesOpt were used in this study, respectively. 

%\begin{center}
%	\url{https://www.ctan.org/pkg/booktabs}
%\end{center}

%\bibliographystyle{unsrt}
%\bibliographystyle{unsrtnat}
\bibliographystyle{abbrvnat}
\bibliography{references}  %%% Remove comment to use the external .bib file (using bibtex).

%\begin{thebibliography}{1}
%
%	\bibitem{Akiba2019}
%	 Takuya Akiba, Shotaro Sano, Toshihiko Yanase, Takeru Ohta, and Masanori Koyama.
%	\newblock Optuna: a next-generation hyperparameter optimization framework.
%	\newblock {\em arXiv:1907.10902}, 2019.
%	
%\end{thebibliography}

\end{document}